\documentclass[onecolumn,nofootinbib]{revtex4}
\usepackage{graphicx}
\usepackage{latexsym}
\usepackage[colorlinks, linkcolor=blue, citecolor=blue, urlcolor=blue]{hyperref}

\begin{document}

\title{New Limits on Coupled Dark Energy from Planck}

\author{Jun-Qing Xia}

\affiliation{Key Laboratory of Particle Astrophysics, Institute of High Energy Physics, Chinese Academy of Science, P. O. Box 918-3, Beijing 100049, P. R. China}

\date{\today}

\begin{abstract}

Recently, the Planck collaboration has released the first cosmological papers providing the high resolution, full sky, maps of the cosmic microwave background (CMB) temperature anisotropies. It is crucial to understand that whether the accelerating expansion of our universe at present is driven by an unknown energy component (Dark Energy) or a modification to general relativity (Modified Gravity). In this paper we study the coupled dark energy models, in which the quintessence scalar field nontrivially couples to the cold dark matter, with the strength parameter of interaction $\beta$. Using the Planck data alone, we obtain that the strength of interaction between dark sectors is constrained as $\beta < 0.102$ at $95\%$ confidence level, which is tighter than that from the WMAP9 data alone. Combining the Planck data with other probes, like the Baryon Acoustic Oscillation (BAO), Type-Ia supernovae ``Union2.1 compilation'' and the CMB lensing data from Planck measurement, we find the tight constraint on the strength of interaction $\beta < 0.052$ ($95\%$ C.L.). Interestingly, we also find a non-zero coupling $\beta = 0.078 \pm 0.022$ ($68\%$ C.L.) when we use the Planck, the ``SNLS'' supernovae samples, and the prior on the Hubble constant from the Hubble Space Telescope (HST) together. This evidence for the coupled dark energy models mainly comes from a tension between constraints on the Hubble constant from the Planck measurement and the local direct $H_0$ probes from HST.

\end{abstract}


\maketitle


\section{Introduction}\label{int}

Current cosmological observations, such as the cosmic microwave background (CMB) measurements of temperature anisotropies and polarization at high redshift $z\sim 1090$ and the redshift-distance measurements of supernovae (SNIa) at $z < 2$, have demonstrated that the universe is now undergoing an accelerated phase of expansion. The nature of dark energy, the mysterious power to drive the expansion, is among the biggest problems in modern physics and has been studied widely. The simplest candidate of dark energy is the cosmological constant, whose equation of state (EoS) $w$ always remains $-1$. Although this model is compatible with the current observational data \cite{planck_fit}, it suffers from the well-known fine-tuning and coincidence problems \cite{ccproblem1,ccproblem2,ccproblem3}.

In order to lift these severe problems, many alternative dynamical dark energy models, such as quintessence \cite{quint1,quint2,quint3,quint4}, phantom \cite{phantom}, k-essence \cite{kessence1,kessence2}, and quintom \cite{quintom1,quintom2,cai}, have been proposed. Interestingly, the dynamical dark energy component is naturally expected to interact with the other components, such as the cold dark matter \cite{copeland98,amendola00} or massive neutrinos \cite{gu03,fardon04}, in the field theory framework. If these interactions really exist, it would open up the possibility of detecting the dark energy non-gravitationally.

In the coupled dark energy models, the quintessence scalar field could nontrivially couple to the cold dark matter component. The presence of the interaction clearly modifies the cosmological background evolutions. The evolution of cold dark matter energy density is dependent on the quintessence scalar field \cite{cde_xia}. The energy density can be transferred between the cold dark matter and the dark energy. On the other hand, the interaction between dark sectors will also affect the evolution of cosmological perturbations (see ref. \cite{cde_xia}, and references therein). The non-zero coupling could shift the matter-radiation equality scale factor, and affect the locations and amplitudes of acoustic peaks of CMB temperature anisotropies and the turnover scales of large scale structure (LSS) matter power spectrum. Furthermore, the coupling affects the dynamics of the gravitational potential, and also affect the late integrated Sachs-Wolfe (ISW) effect \cite{isw}. Therefore, it is of great interest to investigate the non-minimally coupled dark energy models from the current observational data, such as the CMB measurements \cite{Amendola:2002bs,Amendola:2003eq,Bean:2008ac,LaVacca:2009yp,He:2010im,Pettorino:2012ts,Salvatelli2013,Pettorino:2013oxa}, the LSS clustering \cite{Baldi:2010ks,Baldi:2010td}, the weak lensing \cite{DeBernardis:2011iw,Amendola:2011ie}, the cross-correlation between CMB and LSS \cite{cde_xia,Mainini:2010ng,Bertacca:2011in}, and the low redshift observations \cite{Honorez:2010rr}.

Since the Planck collaboration has released the first cosmological papers providing the high resolution, full sky, CMB maps \cite{planck_map}, it is important to study the coupled dark energy models and revisit the constraint on parameters from the latest cosmological probes. In this paper we investigate this kind of model and present the tight constraints from the latest Planck and WMAP9 data, the baryon acoustic oscillations (BAO) measurements from several large scale structure (LSS) surveys, the ``Union2.1'' compilation which includes 580 supernovae, and the CMB lensing data from the Planck measurement. More interestingly, we find a non-zero value for the interaction parameter from the ``SNLS'' supernovae sample and the direct measurement on the Hubble constant from the Hubble Space Telescope (HST). The structure of the paper is as follows: in Sec.II we show the basic equations of background evolution and linear perturbations of the coupled dark energy model. In Sec.III we present the current observational datasets we used. Sec.IV contains our main global fitting results from the current observations, while Sec.V is dedicated to the summary.

\section{Coupled Dark Energy Model}\label{theory}

In this section we briefly review the basic equations for the coupled dark energy model. We refer the reader to refs. \cite{cde_xia,Pettorino:2012ts,Pourtsidou:2013nha} for a very detailed description of all equations involved and effects on the CMB and LSS measurements.

We assume a flat universe described by the Friedmann-Robertson-Walker metric and the adiabatic initial conditions for all components in our analyses. When including the interactions, the conservation of energy momentum for each component becomes \cite{brookfield08}:
\begin{equation}
T^{\mu}_{\nu;\mu}=\beta\phi_{,\nu}T^{\alpha}_{\alpha}~,\label{tmunu}
\end{equation}
where $T^{\mu}_{\nu}$ is the energy-momentum tensors and $\phi$ is the quintessence scalar field. In our analysis we only consider the interaction between cold dark matter and dark energy, the energy conservation equations of cold dark matter and dark energy will be violated:
\begin{eqnarray}
\dot{\rho}^{}_{\rm c}+3\mathcal{H}{\rho}^{}_{\rm c}&=&Q~,\\
\dot{\rho}^{}_{\phi}+3\mathcal{H}{\rho}^{}_{\phi}(1+w^{}_{\phi})&=&-Q~,\label{decoveq}
\end{eqnarray}
where $\rho_{\rm c}$ and $\rho_\phi$ denote the energy densities of cold dark matter and dark energy, $\mathcal{H}=\dot{a}/a$ (the dot refers to the derivative with respect to the conformal time $\eta$) and $Q$ is the interaction energy exchange. Here, we consider the exponential form $\rho^{}_{\rm c}(\phi)=\rho^{\ast}_{\rm c}e^{\beta\phi}$ as the interaction form between quintessence and cold dark matter, where $\rho^{\ast}_{\rm c}$ is the bare energy density of cold dark matter and $\beta$ is the strength of interaction. Then, the energy exchange $Q$ can be written as \cite{cde_xia,Pourtsidou:2013nha}:
\begin{equation}
Q = \beta\dot{\phi}\rho_{\rm c}~.
\end{equation}
When $Q<0$ (or $Q>0$), the energy of cold dark matter (or dark energy) transfers to dark energy (or cold dark matter). Equivalently, the quintessence scalar field $\phi$ evolves according to the Klein-Gordon equation:
\begin{equation}
\ddot{\phi}+2\mathcal{H}\dot{\phi}+a^2V'(\phi)=-a^2\beta\rho^{}_{\rm c}~\label{quinKGeq}.
\end{equation}
In this paper we choose the typical exponential form as the quintessence potential: $V(\phi)=V_{0}e^{-\lambda\phi}$, and the prime denotes the derivative with respect to the quintessence scalar field $\phi\,$: $V'(\phi)\equiv\partial V(\phi)/\partial\phi\,$. We modify these background equations of the coupled dark energy model in the CAMB code \cite{camb} to calculate the cosmological distance information for the SNIa and BAO measurements.

In the synchronous gauge, we could calculate the evolution equations for the perturbation of cold dark matter in the linear regime \cite{cde_xia}:
\begin{eqnarray}
\dot{\delta}_{\rm c}&=&-\theta_{\rm c}-\dot{h}/2+\beta\delta\dot{\phi}~,\\
\dot{\theta}_{\rm c}&=&-\mathcal{H}\theta_{\rm c}-\beta\dot{\phi}\theta_{\rm c}+k^2\beta\delta{\phi}~,
\end{eqnarray}
where $\delta_{\rm c}$ and $\theta_{\rm c}$ are the density perturbation and the gradient of the velocity of the cold dark matter, $\delta\phi$ is the perturbation of dark energy scalar field, and $h$ is the usual synchronous gauge metric perturbation. In the presence of interaction, $\theta_{\rm c}$ will evolve to be nonzero, even if its initial value is zero. Therefore, we compute the perturbation equations in an arbitrary synchronous gauge, instead of the cold dark matter rest frame \cite{cde_xia,Bean:2008ac}. On the other hand, we get the perturbed Klein-Gordon equation in the non-minimally coupled system:
\begin{equation}
\delta\ddot{\phi}+2\mathcal{H}\delta\dot{\phi}+k^2\delta\phi+a^2V''\delta\phi+\dot{h}\dot{\phi}/2=-a^2\beta\rho_{\rm c}\delta_{\rm c}~.
\end{equation}
We include these evolution equations for the perturbations of the coupled dark energy model in the CAMB code to compute the theoretical prediction of the CMB temperature power spectrum.

\begin{figure}[t]
\centering
\includegraphics[scale=0.5]{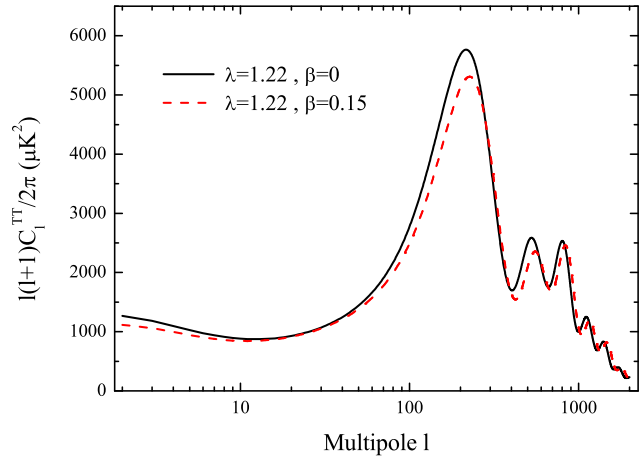}
\caption{The CMB temperature anisotropies for two different models: $\lambda=1.22$, $\beta=0$ (black solid lines) and $\lambda=1.22$, $\beta=0.15$ (red dashed lines).}\label{cmb_eff}
\end{figure}

Here, we use the best fit model of Planck data \cite{planck_like}: $\Omega_bh^2 = 0.02203$, $\Omega_ch^2 = 0.1204$, $\tau = 0.0925$, $h = 0.6704$, $n_s = 0.9619$ and $A_s = 2.215 \times 10^{-9}$ at $k = 0.05$ ${\rm Mpc}^{-1}$. In figure \ref{cmb_eff} we plot the CMB temperature power spectra for two different models: the uncoupled model $\lambda=1.22$, $\beta=0$ (black solid lines) and the coupled system $\lambda=1.22$, $\beta=0.15$ (red dashed lines). We can see that the amplitude on the small scales CMB temperature power spectrum decreases with increasing coupling, due to the earlier epoch of matter-radiation equality $a_{\rm eq}$. And the coupling also shifts locations of the CMB peaks towards smaller scales and suppresses the CMB temperature anisotropies on very large scales (the late-time ISW effect) \cite{cde_xia}.

\section{Data}\label{data}

In our analyses, we consider the following cosmological probes: i) CMB power spectra; ii) the BAO signal in the galaxy power spectra; iii) direct measurement of the current Hubble constant; iv) luminosity distances of type Ia supernovae.

For the Planck data from the 1-year data release \cite{planck_fit}, we use the low-$\ell$ and high-$\ell$ CMB temperature power spectrum data from Planck with the low-$\ell$ WMAP9 polarization data (Planck+WP). We marginalize over the nuisance parameters that model the unresolved foregrounds with wide priors \cite{planck_like}. We also consider the CMB lensing data obtained from Planck \cite{planck_lens} seperately. For comparison, we also use the WMAP9 CMB temperature and polarization power spectra \cite{wmap9} in our calculations.

BAO provides an efficient method for measuring the expansion history by using features in the clustering of galaxies within large scale surveys as a ruler with which to measure the distance-redshift relation. Since the current BAO data are not accurate enough, one can only determine an effective distance \cite{baosdss}:
\begin{equation}
D_V(z)=[(1+z)^2D_A^2(z)cz/H(z)]^{1/3}~.
\end{equation}
Following the Planck analysis \cite{planck_fit}, in this paper we use  the BAO measurement from the 6dF Galaxy Redshift Survey (6dFGRS) at a low redshift ($r_s/D_V (z = 0.106) = 0.336\pm0.015$) \cite{6dfgrs}, and the measurement of the BAO scale based on a re-analysis of the Luminous Red Galaxies (LRG) sample from Sloan Digital Sky Survey (SDSS) Data Release 7 at the median redshift ($r_s/D_V (z = 0.35) = 0.1126\pm0.0022$) \cite{sdssdr7}, and the BAO signal from BOSS CMASS DR9 data at ($r_s/D_V (z = 0.57) = 0.0732\pm0.0012$) \cite{sdssdr9}.

We also add a gaussian prior on the current Hubble constant given by ref. \cite{hst_riess}; $H_0 = 73.8 \pm 2.4$ ${\rm km\,s^{-1}\,Mpc^{-1}}$ (68\% C.L.). The quoted error includes both statistical and systematic errors. This measurement of $H_0$ is obtained from the magnitude-redshift relation of 240 low-z Type Ia supernovae at $z < 0.1$ by the Near Infrared Camera and Multi-Object Spectrometer Camera 2 of HST.

Finally, we include data from Type Ia supernovae, which consists of luminosity distance measurements as a function of redshift, $D_L(z)$. In this paper we consider two SNIa samples: the ``Union2.1'' compilation with 580 samples \cite{union2} and the ``SNLS'' compilation with 473 supernovae reprocessed by ref. \cite{snls}. However, we do not combine them together in the numerical analysis, since we find that these two SNIa samples give quite different constraints on the coupled dark energy model. When calculating the likelihood, we marginalize the nuisance parameters, like the absolute magnitude $M$ and the parameters $\alpha$ and $\beta$, as explained by Ref. \cite{planck_fit}.

\section{Numerical Results}\label{results}

We perform a global fitting of cosmological parameters using the CosmoMC package \cite{cosmomc}, a Markov Chain Monte Carlo (MCMC) code. We assume purely adiabatic initial conditions and neglect the primordial tensor fluctuations. The basic six cosmological parameters are allowed to vary with top-hat priors: the cold dark matter energy density parameter $\Omega_c h^2 \in [0.01, 0.99]$, the baryon energy density parameter $\Omega_b h^2 \in [0.005, 0.1]$, the scalar spectral index $n_s \in [0.5, 1.5]$, the primordial amplitude $\ln[10^{10}A_s] \in [2.7, 4.0]$, the ratio (multiplied by 100) of the sound horizon at decoupling to the angular diameter distance to the last scattering surface $100\Theta_s \in [0.5, 10]$, and the optical depth to re-ionization $\tau \in [0.01, 0.8]$. The pivot scale is set at $k_{s0} = 0.05$ ${\rm Mpc}^{-1}$. There are two more parameters: $\lambda$ and $\beta$ in the potential and coupling forms of the coupled dark energy model.  In addition, CosmoMC imposes a weak prior on the Hubble parameter: $h \in [0.4,1.0]$.

\begin{table}
\caption{The median values and $1\,\sigma$ error bars on some cosmological parameters obtained from different data combinations in the coupled dark energy model. For the interaction strength $\beta$, we quote the $95\%$ upper limits instead.}\label{table1}
\begin{center}

\begin{tabular}{l|c|c|c|c|c|c}

\hline\hline
&WMAP9&Planck+WP&Planck+WP+BAO&Planck+WP+Union&Planck+WP+Lens&Normal Data\\
\hline
$\Omega_bh^2$&$0.02282\pm0.00054$&$0.02196\pm0.00029$&$0.02199\pm0.00026$&$0.02200\pm0.00028$&$0.02210\pm0.00027$&$0.02210\pm0.00025$\\
$\Omega_ch^2$&$0.1066\pm0.0080$&$0.1170\pm0.0044$&$0.1182\pm0.0023$&$0.1159\pm0.0043$&$0.1160\pm0.0037$&$0.1175\pm0.0018$\\
$\Omega_m$&$0.2599\pm0.0616$&$0.3093\pm0.0470$&$0.3097\pm0.0189$&$0.2823\pm0.0336$&$0.3030\pm0.0425$&$0.2997\pm0.0126$\\
$\sigma_8$&$0.8331\pm0.0772$&$0.8410\pm0.0480$&$0.8373\pm0.0262$&$0.8648\pm0.0393$&$0.8337\pm0.0435$&$0.8353\pm0.0172$\\
$H_0$&$72.04\pm7.99$&$67.59\pm4.68$&$67.35\pm1.82$&$70.21\pm3.47$&$67.97\pm4.25$&$68.29\pm1.19$\\
$\beta$&$<0.1446$&$<0.1021$&$<0.0636$&$<0.1002$&$<0.0943$&$<0.0522$\\

\hline\hline
\end{tabular}
\end{center}
\end{table}

\begin{figure}[t]
\centering
\includegraphics[scale=0.45]{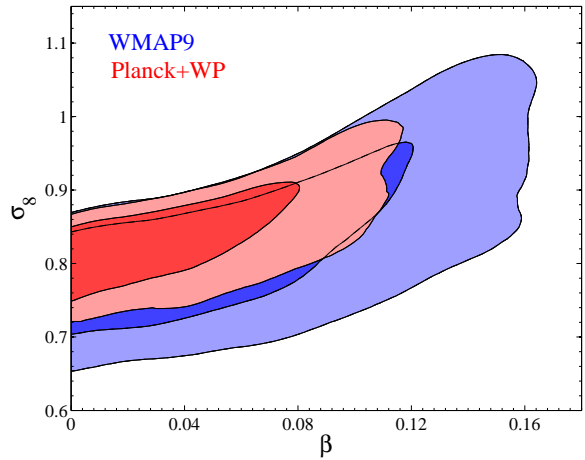}
\caption{Marginalized two-dimensional likelihood (1, $2\,\sigma$ contours) constraints on the parameters $\beta$ and $\sigma_8$ in the coupled dark energy model from the Planck+WP (red) and WMAP9 (blue) data, respectively.}\label{cmb_alone}
\end{figure}

\subsection{Tight Constraints on $\beta$}

Firstly, we consider the constraints on the coupled dark energy model from the Planck+WP and WMAP9 data alone. In table \ref{table1} we list the constraints on some cosmological parameters in the coupled dark energy model from different data combinations. As we know, the CMB anisotropies mainly contain the information about the high-redshift universe, but it is not directly sensitive to phenomena which affect the lower redshift Universe, such as the nature of dark energy. Thus, the WMAP9 data alone can not constrain the parameter $\beta$ of the coupled dark energy model very well. We only obtain the upper limit on the strength of interaction, namely the $95\%$ C.L. constraint is $\beta < 0.145$, which is consistent with previous works \cite{cde_xia,Bean:2008ac}. When we use the more accurate Planck+WP data, the constraint becomes tighter
\begin{equation}
\beta<0.102~~(95\%~{\rm C.L.})~.
\end{equation}
As we mentioned before, the non-minimal coupling shifts the acoustic peaks of CMB temperature anisotropies on the small scales. The small-scale CMB measurements should improve the constraints on the coupling strength. However, since the new Planck data have measured the small-scale CMB power spectrum with very high precision, adding other small-scale CMB data, like Atacama Cosmology Telescope (ACT) \cite{act} and South Pole Telescope (SPT) \cite{spt}, the constraint on $\beta$ is only slightly improved \cite{cde_xia,Pettorino:2013oxa}. In order to save some CPU time for running MCMC, in this work we do not include the ACT and SPT data, since they include many nuisance parameters which significantly slow down our calculations.

In figure \ref{cmb_alone}, we show the two-dimensional contours in the ($\beta$,$\sigma_8$) panel. We can see that $\beta$ is correlated with the $\sigma_8$. In the coupled dark energy model, a positive coupling between cold dark matter and dark energy leads to a high value of $\sigma_8$. That is because we have an energy transfer from cold dark matter to dark energy, which means that there is more dark matter at early times. In this case, the epoch of matter-radiation equality occurs earlier in the coupled model relative to the uncoupled system. Only the very small scale modes could enter the horizon and grow during the radiation dominated era. The growth of perturbations in the coupled model under consideration is enhanced, small-scale power is increased and the value of $\sigma_8$ is larger \cite{cde_xia,Pourtsidou:2013nha}. Using the Planck+WP data alone, we obtain the limit on $\sigma_8$ today of $\sigma_8=0.841\pm0.048$ ($68\%$ C.L.), which is obviously higher than one obtained in the standard $\Lambda$CDM model: $\sigma_8=0.829\pm0.013$ ($68\%$ C.L.) \cite{planck_fit}.

In the standard flat $\Lambda$CDM framework, the constraint on the Hubble constant $H_0$ is significantly improved by the new Planck data, $H_0=67.4\pm1.4$ ${\rm km\,s^{-1}\,Mpc^{-1}}$ at $68\%$ confidence level \cite{planck_fit}. However, this result is obviously in tension with that measured by various lower-redshift methods, such as the direct $H_0$ probe from HST \cite{hst_riess}. The Planck team argued that the local measurements are more likely to be affected by some unknown systematics, such as the effect of a local underdensity, which might lead to this tension \cite{Amendola2013,Verde2013}. Recently, ref. \cite{xia2013} found that if the unknown systematic is not an issue, this $H_0$ tension actually implies that the Planck data favor the dynamical dark energy model, especially one with the EoS $w<-1$. When the model with $w<-1$ is allowed in the analysis, the constraint on $H_0$ from Planck+WP data is consistent with the HST $H_0$ prior. However, this $H_0$ tension cannot be eased in the coupled dark energy model. Planck+WP data yield the $68\%$ C.L. constraint on the Hubble constant of $H_0=67.6\pm4.7$ ${\rm km\,s^{-1}\,Mpc^{-1}}$, which is still significantly lower than the HST measurement. This is because we use the quintensense dark energy model to couple with the cold dark matter. The effective equation of state of dark energy sector cannot be smaller than $-1$ \cite{cde_xia}, therefore, the tension between constraints on $H_0$ still exists. Different from the quintessence scalar field, there are some coupling models where the dark energy component is modeled as a fluid with constant equation of state parameter $w$. It would be very interesting whether the $H_0$ tension can be relaxed when $w$ allowed to be smaller than $-1$ in these models \cite{xia_new}.

\begin{figure}[t]
\centering
\includegraphics[scale=0.35]{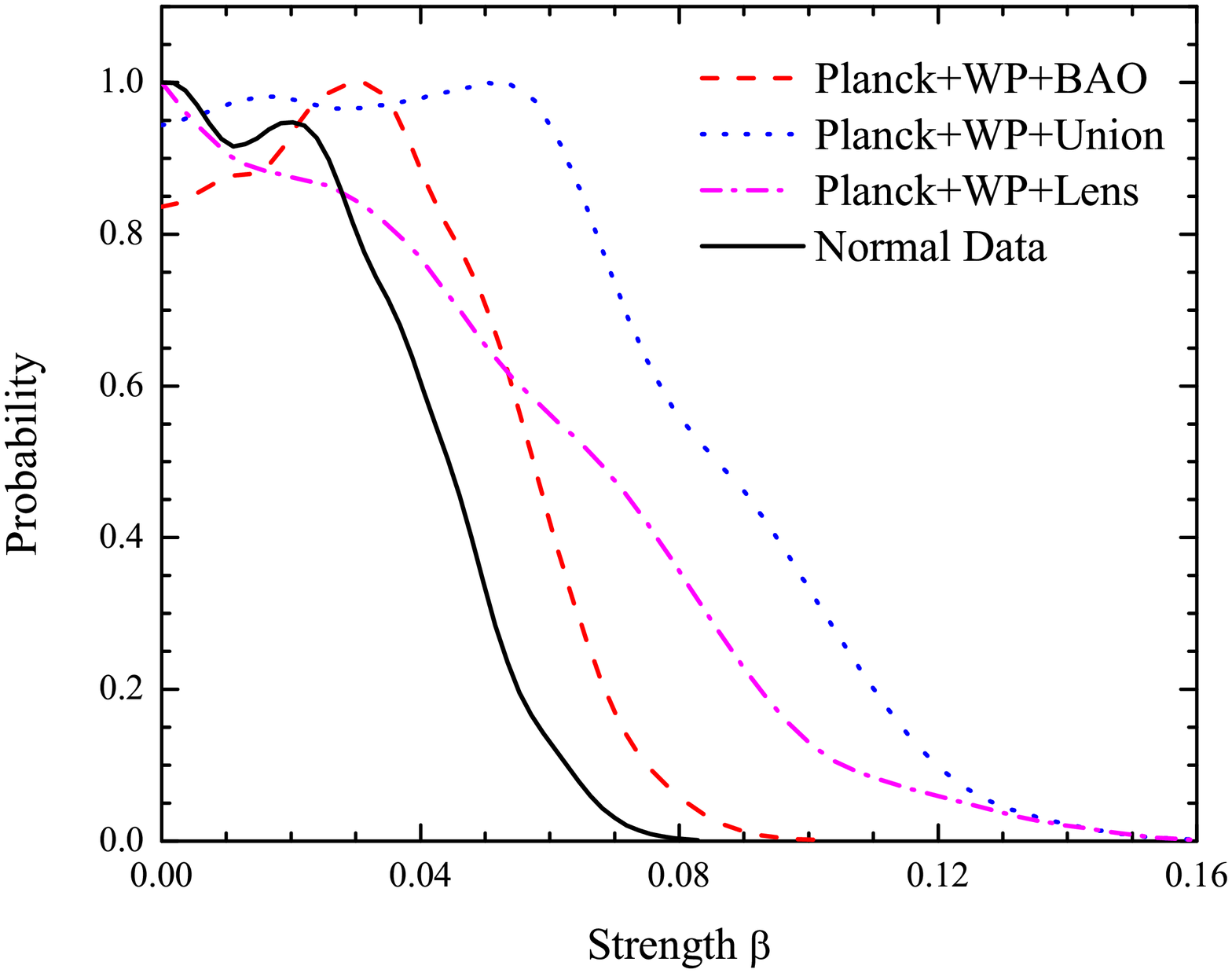}
\caption{One-dimensional posterior distributions of the strength of interaction $\beta$ from various data combinations: Planck+WP+BAO (red dashed line), Planck+WP+Union (blue dotted line), Planck+WP+Lens (magenta dash-dotted line), all ``Normal Data'' together (black solid line).}\label{beta_1d}
\end{figure}

Since the CMB data alone cannot very well constrain the strength of coupling $\beta$ in the coupled dark energy model, we need to add some extra information from the low-redshift probes to break the degeneracy. In our calculations, we consider three kinds of measurements: the BAO signal, the ``Union2.1'' compilation of SNIa and the CMB lensing data from Planck measurement. We also combine these datasets together with the Planck+WP data and call this combined dataset the ``Normal'' data. In figure \ref{beta_1d} we show the one-dimensional posterior distributions of $\beta$ from different data combinations.

The BAO measurements give the tight constraint on the matter density $\Omega_m$. When using Planck+WP and BAO data, the constraint on the strength of coupling is significantly improved, namely $\beta<0.064$ at $95\%$ confidence level. However, the constraining powers of ``Union2.1'' compilation of SNIa and the CMB lensing data are not strong enough for the coupled dark energy model. Planck+WP+Union and Planck+WP+Lens data yield the $95\%$ upper limits on the strength of $\beta<0.100$ and $\beta<0.094$, respectively. When we combine these low redshift probes together, the ``Normal'' data give very stringent upper limit on the strength of interaction between cold dark matter and dark energy:
\begin{equation}
\beta<0.052~~(95\%~{\rm C.L.})~.
\end{equation}
The minimally coupled system is still consistent with these observational datasets. We do not find the evidence for the non-minimally coupled dark energy model for these data combinations.

\subsection{Non-zero Coupling}

\begin{table}
\caption{The median values and $1\,\sigma$ error bars on some cosmological parameters obtained from the Planck+WP, HST and SNLS data in the coupled dark energy model.}\label{table2}
\begin{center}

\begin{tabular}{l|c|c|c|c}

\hline\hline
&Planck+WP&Planck+WP+HST&Planck+WP+SNLS&Tension Data\\
\hline
$\Omega_bh^2$&$0.02196\pm0.00029$&$0.02202\pm0.00028$&$0.02200\pm0.00028$&$0.02203\pm0.00028$\\
$\Omega_ch^2$&$0.1170\pm0.0044$&$0.1125\pm0.0037$&$0.1117\pm0.0049$&$0.1112\pm0.0033$\\
$\Omega_m$&$0.3093\pm0.0470$&$0.2533\pm0.0270$&$0.2453\pm0.0371$&$0.2398\pm0.0229$\\
$\sigma_8$&$0.8410\pm0.0480$&$0.8919\pm0.385$&$0.9031\pm0.0523$&$0.9055\pm0.0385$\\
$H_0$&$67.59\pm4.68$&$73.13\pm3.07$&$74.38\pm4.56$&$74.75\pm2.75$\\
$\beta$&$<0.1021$&$0.0728\pm0.0265$&$0.0735\pm0.0311$&$0.0782\pm0.0217$\\

\hline\hline
\end{tabular}
\end{center}
\end{table}

\begin{figure}[t]
\centering
\includegraphics[scale=0.3]{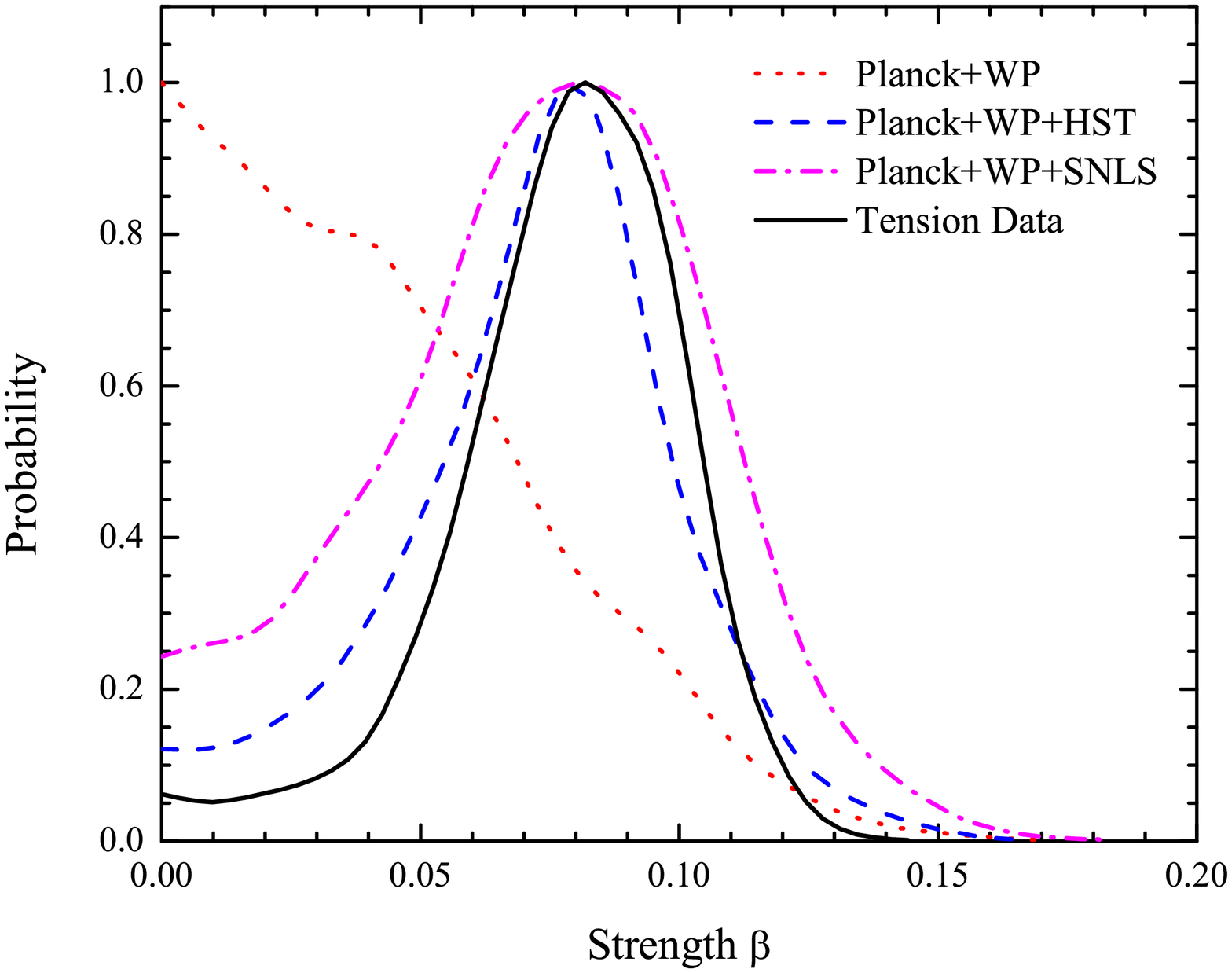}
\includegraphics[scale=0.3]{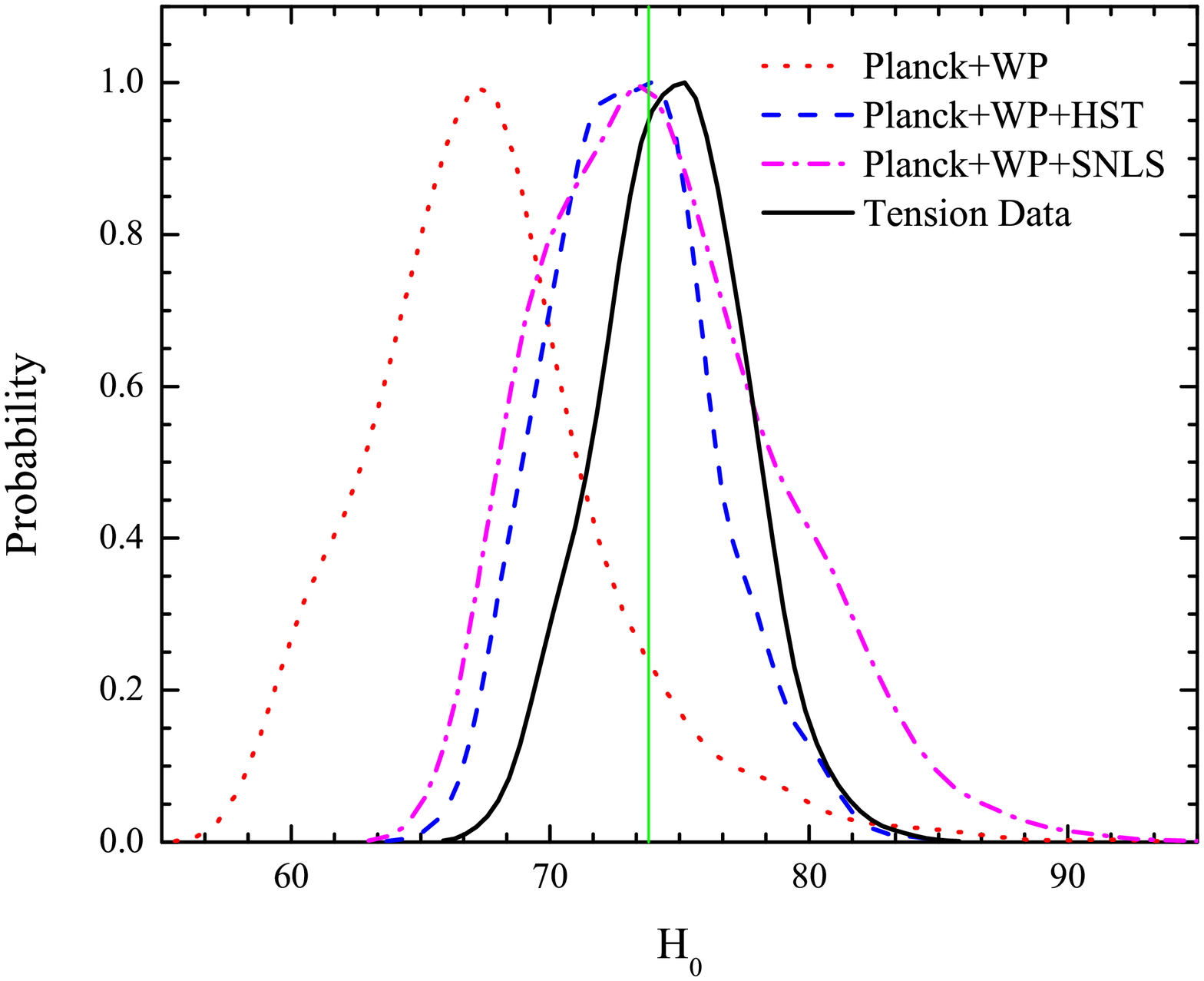}
\caption{One-dimensional posterior distributions of the strength of interaction $\beta$ and the Hubble constant $H_0$ from various data combinations: Planck+WP (red dotted line), Planck+WP+HST (blue dashed line), Planck+WP+SNLS (magenta dash-dotted line), all ``Tension Data'' together (black solid line). The vertical blue dashed line in the right panel denotes the HST prior on the Hubble constant.}\label{tension_1d}
\end{figure}

Besides the ``Normal'' data mentioned above, we also have some other low redshift measurements, like the direct probe on $H_0$ from HST and the ``SNLS'' compilation of SNIa. Interestingly, these two datasets strongly favor a high value of the Hubble constant, which is significantly different from that obtained from the ``Normal'' data \cite{xia2013}. Therefore, we combine these two datasets together to constrain the coupled dark energy model and call it the ``Tension'' data. In table \ref{table2} we list the constraints on some parameters from the ``Tension'' data combinations.

The interaction between the quintessence scalar field and the cold dark matter significantly modifies the evolutions of their energy density \cite{cde_xia}. For a positive coupling, the energy of cold dark matter transfers to dark energy. Consequently, the effective EoS of dark energy becomes smaller than that in the uncoupled system. The current Hubble constant must be increased correspondingly in order to produce the same expansion rate. Therefore, the strength of coupling $\beta$ is correlated with the Hubble constant $H_0$. Including the measurements on $H_0$ could be helpful to break the degeneracy. So we first add the HST gaussian prior on $H_0$ in the calculations. Due to the large discrepancy on $H_0$ between Planck+WP and HST prior, adding the HST prior to the Planck+WP data forces the obtained median value of $H_0$ towards to the higher one, namely $H_0=73.1\pm3.1$ ${\rm km\,s^{-1}\,Mpc^{-1}}$ ($68\%$ C.L.). As a consequence, the degeneracy between $\beta$ and $H_0$ induces a likelihood peak for the strength of coupling:
\begin{equation}
\beta = 0.0728\pm0.0265~~(68\%~{\rm C.L.})~,
\end{equation}
which apparently departs from the minimally coupled system, which is shown in figure \ref{tension_1d} (blue dashed lines).

\begin{figure}[t]
\centering
\includegraphics[scale=0.45]{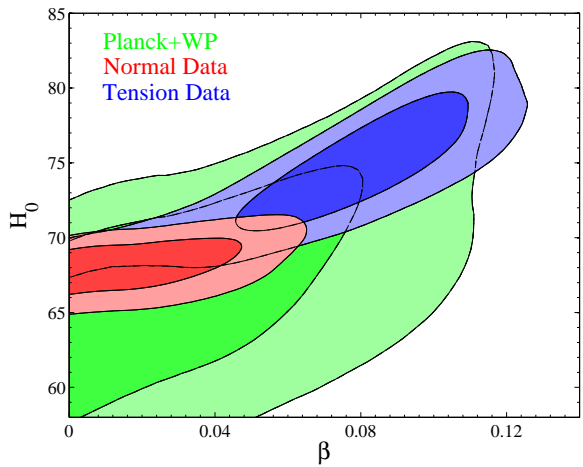}
\caption{Marginalized two-dimensional likelihood (1, $2\,\sigma$ contours) constraints on the parameters $\beta$ and $H_0$ in the coupled dark energy model from Planck+WP alone (green), ``Normal'' data (red) and ``Tension'' data (blue), respectively.}\label{tension_2d}
\end{figure}

We also include the ``SNLS'' supernovae sample into the calculations and find the similar conclusion with that obtained by adding the HST prior. Planck+WP+SNLS data favor a high value of Hubble constant $H_0=74.4\pm4.6$ ${\rm km\,s^{-1}\,Mpc^{-1}}$ ($68\%$ C.L.) and a non-zero coupling parameter $\beta=0.0735\pm0.0311$ ($68\%$ C.L.). Finally, we combine the SNLS and the HST prior together and the situation becomes worse, shown in figure \ref{tension_1d} (black solid lines). The ``Tension'' data combination yields the tight constraints on the Hubble constant $H_0=74.7\pm2.8$ ${\rm km\,s^{-1}\,Mpc^{-1}}$ ($68\%$ C.L.) and the strength of coupling parameter:
\begin{equation}
\beta = 0.0782\pm0.0217~~(68\%~{\rm C.L.})~.
\end{equation}
The preference for a non-zero coupling increases. In figure \ref{tension_2d} we show the two-dimensional contours in the ($\beta$,$H_0$) panel from different data combinations, which is clearly shown that $\beta$ and $H_0$ are correlated. When comparing with the contour obtained from Planck+WP data, the ``Normal'' data give consistent constraint on the Hubble constant. Therefore, this data combination only reduces the correlation between $\beta$ and $H_0$ to give better constraint on $\beta$. However, the ``Tension'' data favor a high value of the Hubble constant. Using this data combination shifts the two-dimensional contour towards high values of $H_0$. Consequently, a non-zero coupling parameter is slightly favored by this data. But it is not strong enough to claim a deviation from the standard $\Lambda$CDM model.

\section{Summary}\label{sum}

In this paper we have presented the latest cosmological constraints on the coupled dark energy models, in which the quintessence scalar field non-minimally couples to the cold dark matter, from the recent Planck measurements. Since the CMB anisotropies mainly contain the information about the high-redshift universe, Planck+WP data alone cannot constrain the strength of coupling strongly, namely the $95\%$ C.L. upper limit is $\beta<0.102$.

When we combine the Planck+WP with the BAO, SNIa ``Union2.1'' compilation and the CMB lensing data from Planck, this ``Normal'' data yield a very tight constraint on the strength, $\beta<0.052$ at $95\%$ confidence level. Different from the dark energy model with a constant equation of state parameter $w$, the tension between constraints on $H_0$ from the Planck+WP and the HST $H_0$ prior cannot be eased in the coupled dark energy models, since in these models the effective EoS of dark energy is always larger than $-1$.

Finally, we use the HST prior and the SNIa ``SNLS'' sample to break the degeneracy between $\beta$ and $H_0$. Since these two data strongly favor a high value of the Hubble constant, we find an interesting preference for the non-zero coupling: $\beta = 0.0782\pm0.0217$ ($68\%$ C.L.) from the ``Tension'' data combination. This result mainly comes from a tension between constraints on the Hubble constant from the Planck measurement and the local direct $H_0$ probes and needs to be clarified further.


\section*{Acknowledgements}

We acknowledge the use of the Legacy Archive for Microwave Background Data Analysis (LAMBDA). Support for LAMBDA is provided by the NASA Office of Space Science. JX is supported by the National Youth Thousand Talents Program and the grants No. Y25155E0U1 and No. Y3291740S3.


\end{document}